# Comb-rooted multi-channel synthesis of ultra-narrow optical frequencies of few Hz linewidth

Heesuk Jang[1,3], Byung Soo Kim[1,3], Byung Jae Chun[1,2], Hyun Jay Kang[1], Yoon-Soo Jang[1], Yong Woo Kim[1], Young-Jin Kim[1,2,*] and Seung-Woo Kim[1,*]

[1]Department of Mechanical Engineering, Korea Advanced Institute of Science and Technology (KAIST), 291 Daehak-ro, Yuseong-gu, Daejeon 34141, Republic of Korea

[2]School of Mechanical and Aerospace Engineering, Nanyang Technological University (NTU), 50 Nanyang Avenue, Singapore, 639798, Singapore.

[3]These authors equally contributed to this work.

*Corresponding authors: *swk@kaist.ac.kr, yj.kim@ntu.edu.sg*

**We report a multi-channel optical frequency synthesizer developed to generate extremely stable continuous wave lasers directly out of the optical comb of an Er-doped fiber oscillator. Being stabilized to a high-finesse cavity with a fractional frequency stability of $3.8 \times 10^{-15}$ at 0.1 s, the comb-rooted synthesizer produces multiple optical frequencies of ultra-narrow linewidth of 1.0 Hz at 1 s concurrently with an output power of tens of mW per each channel. Diode-based stimulated emission by injection locking is a key mechanism that allows comb frequency modes to sprout up with sufficient power amplification but no loss of original comb frequency stability. Channel frequencies are individually selectable with a 0.1 GHz increment over the**

**entire comb bandwidth spanning 4.25 THz around a 1550 nm center wavelength. A series of out-of-loop test results is discussed to demonstrate that the synthesizer is able to provide stable optical frequencies with the potential for advancing diverse ultra-precision applications such as optical clocks comparison, atomic line spectroscopy, photonic microwaves generation, and coherent optical telecommunications.**

Narrow spectral linewidth lasers with wavelength tunability are needed in various areas of science and technology. Especially, ultra-stable lasers capable of providing extremely narrow linewidths of few Hz or below play important roles to make fundamental scientific breakthroughs. The examples include the optical clock technology[1], clock-based geodesy [2], physical constants measurement [3], and general relativity experiments[4]. Such narrow linewidths are also demanded for industrial uses such as precision spectroscopy[5], low noise microwaves generation[6], and coherent optical telecommunications. Currently, as depicted in Fig. 1A, semiconductor lasers are available with linewidths narrowed down to few hundreds of kHz by imbedding distributed feedback (DFB) or distributed Bragg reflector (DBR) diodes[7]. Incorporating an external cavity leads to further linewidth reduction to few kHz while the wavelength tunability extends to hundreds of nanometers[8]. Meanwhile, even though their wavelength tunability is not comparable to those of semiconductor lasers, fiber-based lasers containing narrow bandwidth filters are also offering narrow linewidths of few kHz, and extreme linewidth reduction even to tens of hertz can be achieved by employing optical feedback such as stimulated Brillouin scattering, Rayleigh backscattering and self-injection[9].

The frequency stability and accuracy of a tunable laser can be enhanced by coherent phase-locking to an optical comb that acts as a frequency ruler over an extensive wavelength range [10]. This comb-based optical frequency stabilization has been demonstrated with reference to optical references[11] or microwave references[12]. Besides, with the aid of passive injection locking, extracting a single frequency mode directly from an optical comb has been put into practice as an effective way of synthesizing continuous wave lasers[13,14]. Effort is also being made for direct synthesis of arbitrary optical frequencies by utilizing micro-cavity combs as a means of flexible frequency up-conversion chain[15]. Here, we propose a multichannel optical frequency synthesizer capable of generating ultra-narrow lasers directly out of the optical comb of a fiber oscillator. The focus is placed on the diode-based stimulated emission mechanism by injection locking specially devised to allow original frequency modes to sprout up through slave diode lasers with tens of mW power while maintaining the superb comb frequency stability. This comb-rooted optical frequency synthesizer, being self-stabilized to a high-finesse cavity, is able to produce ultra-narrow lasers of 1.0 Hz linewidth with tens of mW per each. The wavelength selectivity reaches a 34 nm (4 THz) range around a 1550 nm centre wavelength with 0.1 GHz increment.

**Results and Discussions**

**Comb-rooted optical frequency synthesis.** Figure 1B illustrates the concept of multi-channel optical synthesizer pursued in this study to produce highly stable optical frequencies with ultra-narrow linewidth over a broad spectral range. The main idea is to extract multiple frequency modes from the optical comb of a mode-locked fiber oscillator at different frequency locations concurrently with sufficient power amplification. In addition, one of extracted comb frequency modes - designated $\nu_{ref}$ - is locked to the resonance peak of a high-finesse cavity. At the same time, the carrier-envelop offset frequency $f_o$, being detected through a self-referencing *f-2f* interferometer, is regulated to prevent phase noise propagation across frequency modes. This comb stabilization permits the optical comb to maintain an ultra-high level of stabilization inherited from the high-finesse cavity, thereby allowing all the extracted comb modes to turn to stable continuous-wave lasers of few Hz narrow linewidth. This scheme requires no extra lasers and further the long-term stability may be procured by locking an extracted frequency mode additionally to an appropriate optical clock signal. Moreover, stable microwave signals can readily be synthesized with flexibility by pairing two comb modes of selected optical frequency difference.

**Frequency mode extraction.** Figure 1C describes how original frequency modes are extracted individually and amplified to be used as stable, robust continuous wave lasers. The optical comb used here as the frequency ruler is produced from an Er-doped fiber femtosecond laser (C-fiber, Menlosystems GmbH). The optical comb consists of ~ $10^5$ frequency modes evenly distributed with a 100 MHz spacing in the spectral domain over a 75 nm spectral bandwidth around a 1550 nm center wavelength. The total comb power is 20 mW while a single frequency mode is allocated merely a tiny power of ~ 200 nW. Any frequency mode within the optical comb may be selected for extraction in parallel or sequentially for multi-channel synthesis. For each selected mode, a band-pass filter unit is devised by combining a fiber Fabry-Perot (FFP) filter with a fiber Bragg grating (FBG). The FFP filter is set to provide multiple transmission windows equally spaced, with each window being narrow enough to let only a single frequency mode pass through. The FBG filter is arranged to offer a single large transmission window in overlap with three FFP windows, with the target frequency mode being contained within the FFP window. The optical spectrum observed after the FBG filter consequently consists of the target frequency mode in the middle and two weak modes on its both sides (Further details are given in Methods).

**Power amplification.** Diode-based stimulated emission by injection locking[16] is used to amplify the extracted frequency mode. Specifically, the extracted frequency mode is injected as the seed laser to a distributed feedback (DFB) laser diode, of which the operating range covers the whole spectral bandwidth of the optical comb. Stimulated emission occurring inside the DFB laser diode boosts the optical power of the seed laser while sustaining the superb linewidth and absolute frequency position of the original frequency mode[14]. Besides, the injection locking mechanism selectively amplifies only the target frequency mode of strongest optical power, whereas weak two side modes are rejected by means of mode competition. This all-fiber-based injection locking technique is able to provide an amplification factor of 40 – 50 dB with neither frequency shifting nor linewidth broadening. In consequence, the amplified target frequency mode exhibits a very high signal-to-noise suppression ratio of ~ 50 dB, thereby evolving to an independent continuous wave laser of ~ 1.0 Hz linewidth with tens of mW optical power.

**Comb stabilization**. Figure 2 illustrates how the optical comb used as the frequency ruler in the synthesizer is stabilized to a high-finesse optical cavity directly without extra lasers. In the first place, in control of the repetition rate $f_r$, a reference laser line is produced at a preassigned position at $\nu_{ref}$ and subsequently locked to its nearest resonance peak of the high-finesse cavity by applying the Pound-Drever-Hall (PDH) technique[17]. Simultaneously, the carrier-envelop offset frequency $f_o$ detected using an *f-2f* interferometer is nullified to achieve a zero carrier-envelop offset[18]. In consequence, any frequency mode $\nu_{opt}$ becomes positioned at an exactly integer multiple of the repetition rate, i.e. $\nu_{opt} = n \times f_r$. The transmission curve of the high-finesse cavity (Fig. 2B) has a narrow profile of 8 kHz bandwidth (FWHM) corresponding to a high finesse value of 400,000. The PDH control signal (Fig. 2C) monitored during the PDH stabilization exhibits a good linear sensitivity of 75 µV/Hz, permitting the frequency mode $\nu_{ref}$ to be phase-locked to the high-finesse cavity with a sub-Hz control resolution. The PDH control signal (Fig. 2D) digitized at a sampling rate of 500 kS/s over a time interval of 20 s leads to an Allan deviation of $1.0 \times 10^{-12}$ at $10^{-5}$ s or $3.0 \times 10^{-15}$ at 1 s averaging (Fig. 2E). The histogram (Fig. 2F) obtained by projecting the PDH control signal[19], over a 20 s period with consecutive 1 s averaging steps in this particular case, indicates that the instantaneous linewidth of the frequency mode $\nu_{ref}$ reaches 0.20 Hz FWHM.

**Arbitrary optical frequency synthesis.** Figure 3 illustrates how an arbitrary frequency $\nu_t$ is synthesized as $\nu_t = \nu_{ref} + m \times f_r$ with the position index $m$ being redefined relative to the comb mode $\nu_{ref}$ being locked to the high-finesse cavity. The first step is to select an FFP filter of which the transmission window embraces the target frequency $\nu_t$ with reference to a pre-calibrated relation between the position index $m$ and the relative position of each filter within the given FFP array. The selected FFP filter is fine-tuned by control of its temperature to let the target frequency $\nu_t$ pass through the transmission window. The second step is to adjust the DFB laser diode by control of input current so that the target frequency $\nu_t$ is well placed inside the injection locking range. This two-step adjustment procedure is repeated when the target frequency has to be moved from position to another. Fig. 3B shows three channel frequencies (Fig. 3B) synthesized in sequence for $m$, $m+1$, $m+2$, with $m=4269$ corresponding to an optical frequency of 196,078,431.000 MHz. The synthesized optical frequencies were analyzed by beating them with an auxiliary optical comb that was stabilized to the same high-finesse cavity as described in Fig.2. The resulting RF beat notes revealed that all the three frequencies are precisely positioned at their individual target absolute positions with a nominal deviation of 1.0 Hz with a good spectral stability (Fig. 3C).

**Comb-rooted multi-wavelength synthesis.** The performance of the comb-rooted synthesizer was evaluated in terms of the absolute frequency position, spectral linewidth, phase noise and long-term stability through a series of out-of-loop measurements as presented in Fig. 4. For the sake of absolute testing, two separate synthesizers - designated Comb #1 and Comb #2 - were prepared to concurrently produce three frequency channels per each at 1530 nm, 1555 nm, and 1564 nm (Figs. 4B & 4C). The linewidth of each channel was evaluated individually by heterodyning one frequency ν of Comb #1 with its counterpart ν´ of Comb #2. The resulting RF beat of $f_{1564} = \nu_{1564} - \nu´_{1564}$ appeared to have a Lorentzian profile, of which the linewidth was measured to be 1.50 Hz (FWHM) for a 1.8 s acquisition time. Assuming that both the frequencies of $\nu_{1564}$ and $\nu´_{1564}$ equally contribute to the temporal fluctuation of $f_{1564}$, the linewidth of $\nu_{1564}$ (or $\nu´_{1564}$) was estimated to be 1.0 Hz by dividing the linewidth of $f_{1564}$ by $\sqrt{2}$. Other two channels at 1555 nm and 1530 nm were assessed in the same way, of which linewidths were found almost the same as that of $\nu_{1564}$. This out-of-loop teat validates that the comb-rooted synthesizer, either Comb #1 or Comb #2, is able to generate ultra-narrow 1.0 Hz linewidth over a wavelength range of 34 nm. Next, the whole optical spectrum of the three channels of Comb #1 was monitored continuously over a 30 minute period, of which the result exhibited a reliable long-term operation with all the channels maintaining a signal-to-noise ratio (SNR) of 60 dB (Fig. 4D). Lastly, the phase noise (Fig. 4E) of the beat $f_{1564} = \nu_{1564} - \nu'_{1564}$ analyzed using an RF spectrum analyzer showed -42.5 dBc/Hz at 10 Hz along with a slight servo bump at a 100 kHz offset that is attributable to the control bandwidth of the extra-cavity AOM used for comb stabilization (Fig. 2). The Allan deviation (Fig. 4F) calculated by the phase noise frequency-to-time conversion[20] revealed a high level of signal repeatability of $3.8\times10^{-15}$ at 0.1 s averaging.

**Conclusions**

To conclude, the comb-rooted synthesizer configured in this study has been proven to be able to generate optical frequencies of extremely narrow linewidth of 1.0 Hz at 1 s with a high stability of $3.8\times10^{-15}$ at 0.1 s. Diode-based stimulated emission by injection locking plays the essential role of permitting concurrent extraction of multiple frequency modes with an output power of tens of mW per each channel. Channel frequencies are selectable with a 0.1 GHz increment over a broad bandwidth of 4.25 THz around a 1550 nm center wavelength. The comb-rooted synthesizer is anticipated to act as a multi-channel optical source for diverse applications such as optical clocks comparison, atomic line spectroscopy, photonic microwaves generation and coherent optical telecommunications.

**Methods**

**High-finesse optical cavity as an optical frequency reference.** The high-finesse cavity (ATF6300, SLS) used in Fig. 1B (and 2A) is made of ultra-low expansion (ULE) glass of a 50 mm length. With a finesse value of 400,000, the cavity provides resonance peaks of 8 kHz bandwidth (FWHM) evenly distributed with a 3.14 GHz free spectral range (FSR) in both the transmitted and reflected light. The input beam to the cavity is delivered through an achromatic lens with precise alignment so as to achieve accurate mode matching with respect to the end mirrors of the cavity. For protection from environmental disturbances, the cavity is installed inside a vacuum chamber of $10^{-7}$ Torr with strict temperature control of ± 0.001 °C at 34 °C. In addition, the vacuum chamber is surrounded with an acoustic enclosure seated on an active vibration-isolation support.

**Frequency mode extraction.** The spectral filter devised in Fig. 1C for frequency mode extraction is a composite system comprising a fiber Fabry-Perot (FFP) filter and a fiber Bragg grating (FBG) filter in sequence. First, the fiber FFP filter offers multiple transmission windows equally distanced with a 50 GHz free spectral range (FSR). Each FFP window has a 100 MHz transmission bandwidth (FWHM) with a 500 finesse value, being narrowed to permit only one mode to pass through. Accordingly, the original comb is sorted to have a widened 50 GHz mode spacing. Second, the fiber FBG filter provides a single transmission opening of a large 100 GHz width (FWHM), embracing three FFP transmission windows symmetrically – one main mode in the center plus two weak modes in both sides. Last, the DFB laser diode attached to the FBG filter is tuned by input current control so that the central main mode is amplified through diode-based stimulated emission by injection locking, whereas the side modes are suppressed by the selective nature of the diode-based injection locking.

**Control strategy for comb stabilization.** Comb stabilization in Fig. 2A is realized by incorporating three control devices; an acousto-optic-modulator (AOM), a piezo-electric transducer (PZT) and a pump current controller[21]. First, the AOM is connected to the output port of the source oscillator, so it can shift the whole optical comb laterally in parallel.[22] Specifically, let $f_{AOM}$ be the instantaneous amount of frequency shift induced by the AOM, then the whole optical comb undergoes a uniform shift as $\nu_{opt} = n \times f_r + f_o - f_{AOM}$ with $n$ being the frequency mode number and $f_r$ the pulse repetition rate. Second, the PZT is set to translate the end mirror of the source oscillator, so it regulates $\nu_{opt}$ by varying the pulse repetition rate $f_r$. Third, the pump current controller reins the carrier-envelop offset frequency $f_o$ by adjusting the input current to the pumping diode of the source oscillator. Now, the control procedure for comb stabilization starts with selecting a suitable frequency mode as $\nu_{ref}$ and positioning it within a resonance peak of the high-finesse cavity. The PDH servo is then set to nullify the PDH error signal by tuning both $f_{AOM}$ and $f_r$ through activation of both the AOM and the PZT with a dual-servo scheme[23]. Note that the PZT installed inside the source oscillator is able to tune $\nu_{ref}$ over a wide dynamic range of ~ 1 GHz by regulating $f_r$ but its control bandwidth is as slow as ~ 20 kHz. On the other hand, the AOM installed outside the source oscillator offers a small dynamic range of 6 MHz but its control bandwidth is as fast as 200 kHz. Hence, the dual-servo scheme combining the PZT with the AOM permits the PDH locking to execute over a large dynamic range of 1 GHz with a fast control bandwidth of 200 kHz. Meanwhile, the carrier-envelop offset frequency $f_o$ is detected using an *f-2f* interferometer with a 40 dB signal-to-noise ratio. The detected $f_o$ is then controlled by activating the pump current controller with a 40 kHz bandwidth so as to counterbalance the AOM frequency, i.e. $f_o - f_{AOM} = 0$. This counterbalance control is intended for the whole optical comb to be stabilized with a zero carrier-envelop offset, so all the frequency modes turn out exactly integer multiples of the pulse repetition rate, i.e. $\nu_{opt} = n \times f_r$, without relying on an external radio-frequency clock of which short-term stability is limited.

**Pound-Drever-Hall (PDH) locking.** In Fig. 2A, the mode $\nu_{ref}$ extracted at a 1564 nm wavelength with 20 mW optical power is made incident to the high-finesse cavity, which is then phase-modulated through an electro-optic modulator (EOM) with a 48 MHz driving frequency. The PDH locking control is conducted to tune the frequency mode $\nu_{ref}$ to precisely match with the designated resonance peak of the high-finesse cavity. A linear polarizer is inserted before the EOM to minimize the spurious effect due to residual amplitude modulation (RAM) by adjusting the polarization of the input beam to be aligned precisely along the main axis of the EOM crystal material. In addition, the EOM is thermo-electrically temperature stabilized to reduce the RAM-related error arising from temperature-dependent birefringence[24]. Further, optical isolators are inserted so as to block the scattered and/or reflected light that would drift the zero level of the PDH error signal owing to the so-called etalon effect caused by imperfect fiber splicing between the EOM and the high-finesse cavity[25].

**Acousto-optic modulator (AOM) for comb stabilization.** In Fig. 2A, the PDH error signal is corrected by tuning the extracted mode $\nu_{ref}$ using a fiber-coupled AOM attached to the outlet of the source oscillator. This extra-cavity AOM operates with a 40 MHz driving frequency, shifting all the modes within the optical comb in parallel with a modulation bandwidth of 200 kHz. The 1st order down-shifted diffraction of the AOM output is used, so the *n*-th frequency mode is spectrally positioned as $\nu_{opt} = n \times f_r + f_o - f_{AOM}$ with $f_{AOM}$ being the instantaneous amount of frequency shift induced by the AOM.

*f-2f* **interferometer.** For the detection of the carrier-envelop offset frequency $f_o$ depicted in Fig. 2A, a portion of the source beam of the source oscillator is diverted to a highly nonlinear fiber (HNLF) for spectral broadening[26]. In order to create an octave-spanning supercontinuum, the input beam to the HNLF is amplified to 200 mW and compressed to 60 fs beforehand. The *f-2f* interferometer configured here is a Mach-Zehnder type with two optical paths; one path takes in original wavelengths and the other receives frequency-doubled wavelengths near 1030 nm through a PPLN (periodically poled lithium niobate) crystal. The temporal overlapping between *f* and *2f* pulses in the time domain is precisely adjusted by tuning the delay between the two optical paths. The $f_o$ signal is monitored using a radio-frequency spectrum analyzer with a 3 kHz resolution bandwidth, which yields a signal-to-noise ratio of 40 dB. In free running state the $f_o$ signal floats over hundreds of kHz, so it is down-converted using a frequency divider to fit in the 40 kHz control bandwidth of the pump current controller[27].

**Phase noise propagation over frequency modes.** In the comb stabilization scheme of Fig. 2A, the frequency mode locked to the high-finesse cavity is selected at the mode number of $\nu_{ref}$ with its optical frequency being $\nu_{ref} = n_{ref} \times f_r + (f_o - f_{AOM})$. Then, the optical frequency of the $n$-th frequency mode is expressed as $\nu_{opt} = n \times f_r + (f_o - f_{AOM})$ or $\nu_{opt} = r \times \nu_{ref} + (1 - r) \times (f_o - f_{AOM})$ with $r$ being defined as the mode number ratio of $r = n / n_{ref}$. The phase noise power spectral density (PSD) of $\nu_{opt}$ is then obtained as $S_{opt}(f) = r^2 S_{ref}(f) + (1 - r)^2 [S_o(f) + S_{AOM}(f)] \approx r^2 S_{ref}(f)$ since both $S_o(f)$ and $S_{AOM}(f)$ may be assumed to be much smaller than $S_{ref}(f)$. The ratio $r$ becomes almost unity for all the frequency modes within an optical comb, so it is concluded that there is no significant propagation of phase noise between $S_{ref}(f)$ and $S_{opt}(f)$, i.e., $S_{opt}(f) \approx S_{ref}(f)$[28].

**Acknowledgements**

This work was supported by the National Research Foundation of the Republic of Korea (NRF-2012R1A3A1050386). Y.-J. Kim acknowledges support from the Singapore National Research Foundation (NRF-NRFF2015-02).

**Author contributions**

The project was planned and overseen by S.-W.K. in collaborations with Y.-J.K. Experiments were performed by H.J. and B.S.K. with supports of B.J.C., H.J.K. and Y.-S.J. All authors contributed to the manuscript preparation.

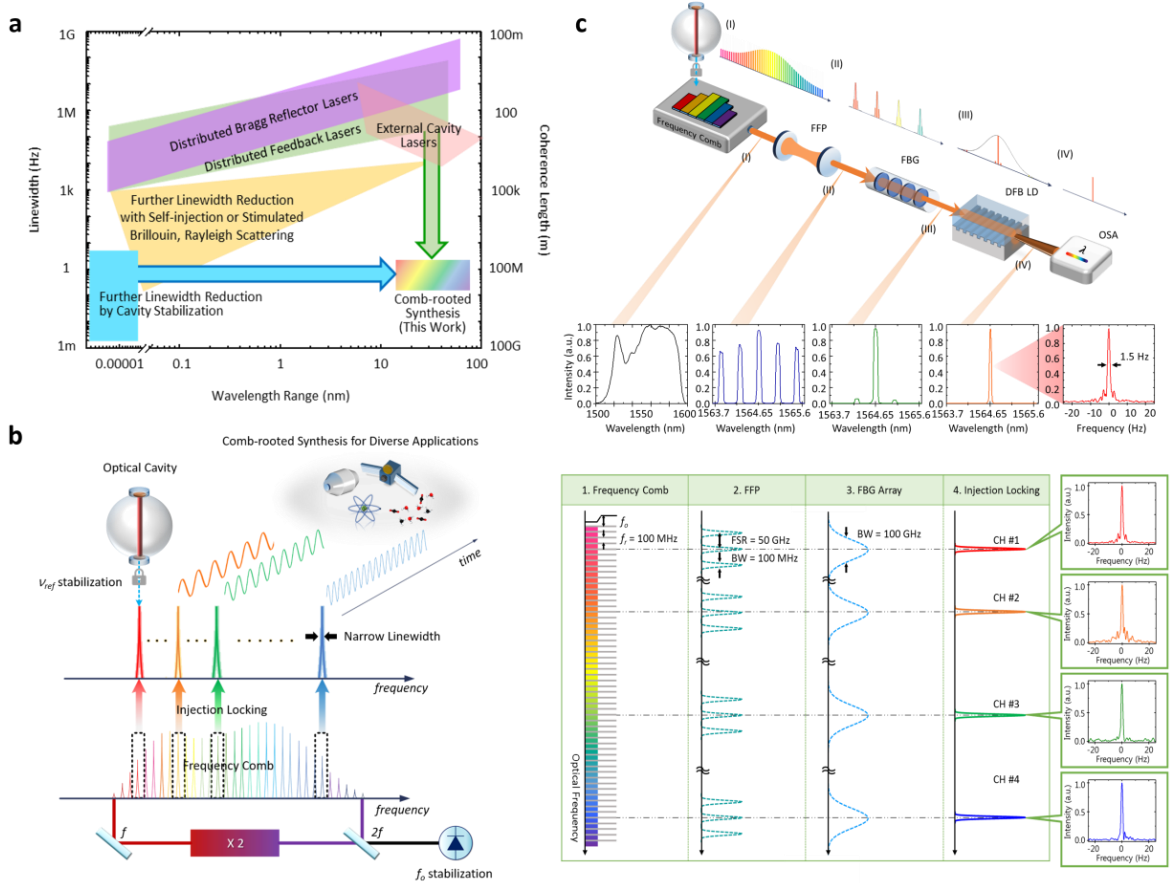

**Fig. 1. Comb-rooted optical frequency synthesis.** (**a**) Spectral linewidth vs. tunable range of state-of-the-art continuous-wave lasers; distributed feedback (DFB) lasers, distributed Bragg reflector (DBR) lasers[7], external cavity laser diodes (ECLDs)[8], linewidth-reduced lasers by optical feedback [9,29] and electrical feedback[30] are considered here in comparison with the comb-rooted synthesizer developed in this study. (**b**) Concept of comb-rooted optical frequency synthesis. Multiple frequency modes are extracted concurrently with subsequent conversion to continuous wave lasers to be used for diverse applications. (**c**) Diode-based stimulated emission by injection locking for power amplification & multi-wavelength comb-rooted synthesis. Abbreviations are; FFP: fiber Fabry-Perot filter, FBG: fiber Bragg grating and OSA: optical spectrum analyzer.

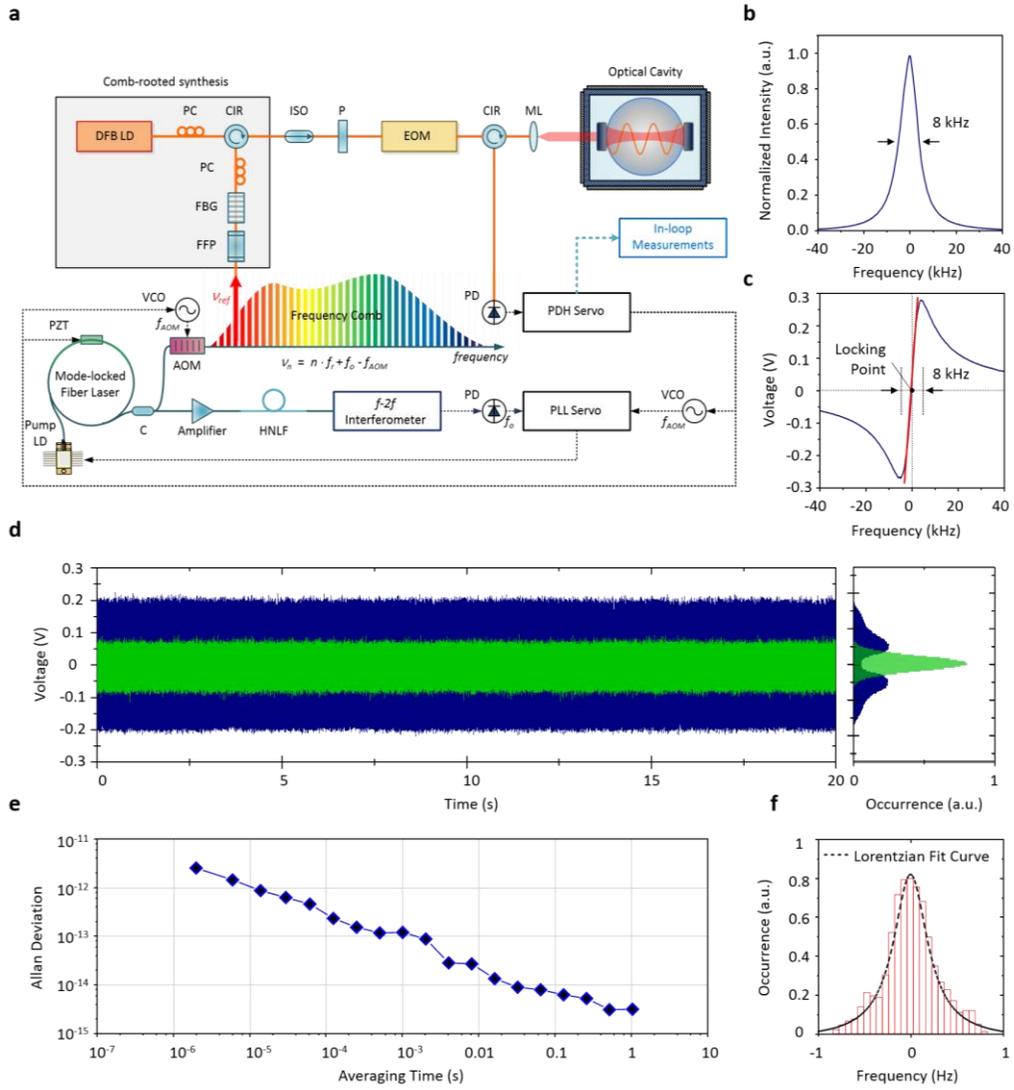

**Fig. 2. Comb stabilization to a high-finesse optical cavity.** (**a**) An extracted frequency mode $\nu_{ref}$ is locked directly to a high-finesse optical cavity by controlling $f_{AOM}$ and $f_r$ using the Pound-Drever-Hall (PDH) method. At the same time, the carrier-envelop offset frequency $f_o$ is detected through an *f-2f* interferometer and regulated to suppress $f_{AOM}$ instantly so as to achieve zero-offset comb stabilization. (**b**) Transmission curve of the high-finesse cavity used for Pound-Drever-Hall (PDH) locking. (**c**) PDH control signal monitored as a function of the deviation of $\nu_{ref}$ from the cavity resonance frequency. (**d**) Histograms of the PDH control signal before (blue) and after (green) comb stabilization (**e**) Allan deviation of the PDH control signal. (**f**) Histogram obtained by averaging the PDH control signal for 1 s. Abbreviations are; PZT: piezo-electric transducer, Pump LD : pump laser diode, AOM: acousto-optic modulator, FFP: fiber Fabry-Perot filter, FBG: fiber Bragg grating, PC: polarization controller, DFB LD: Distributed feedback laser diode, CIR: circulator, ISO: isolator, P: linear polarizer, EOM: electro-optic modulator, ML: mode-matching lens, PD: photodetector, PDH servo: Pound-Drever-Hall servo, C: coupler, HNLF: highly nonlinear fiber, PLL servo: phase-locked loop servo, VCO : voltage controlled oscillator.

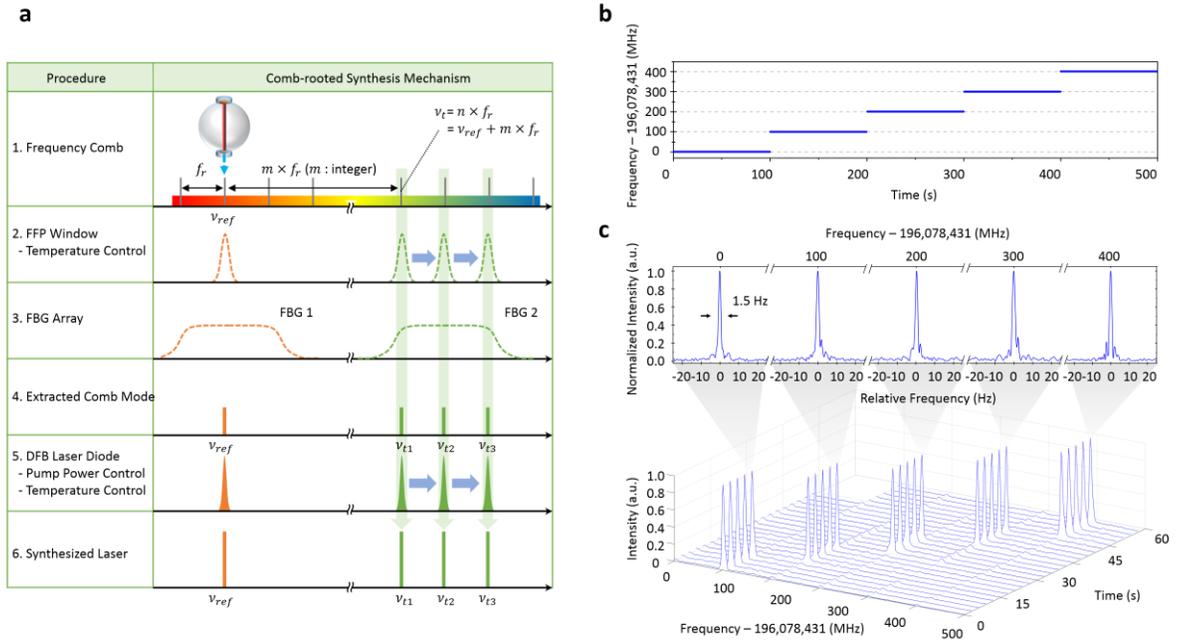

**Fig. 3. Optical frequency synthesis.** (**a**) Optical frequency tuning process. (**b**) A sequence of optical frequencies synthesized with a 100-MHz step increase consecutively from an initial frequency of 196,078,431.000 MHz (gate time: 100 ms). (**c**) Time traces of optical spectral and linewidth profiles (RBW: 1 Hz). Abbreviations are; FFP: fiber Fabry-Perot filter, FBG: fiber Bragg grating, DFB LD: Distributed feedback laser diode.

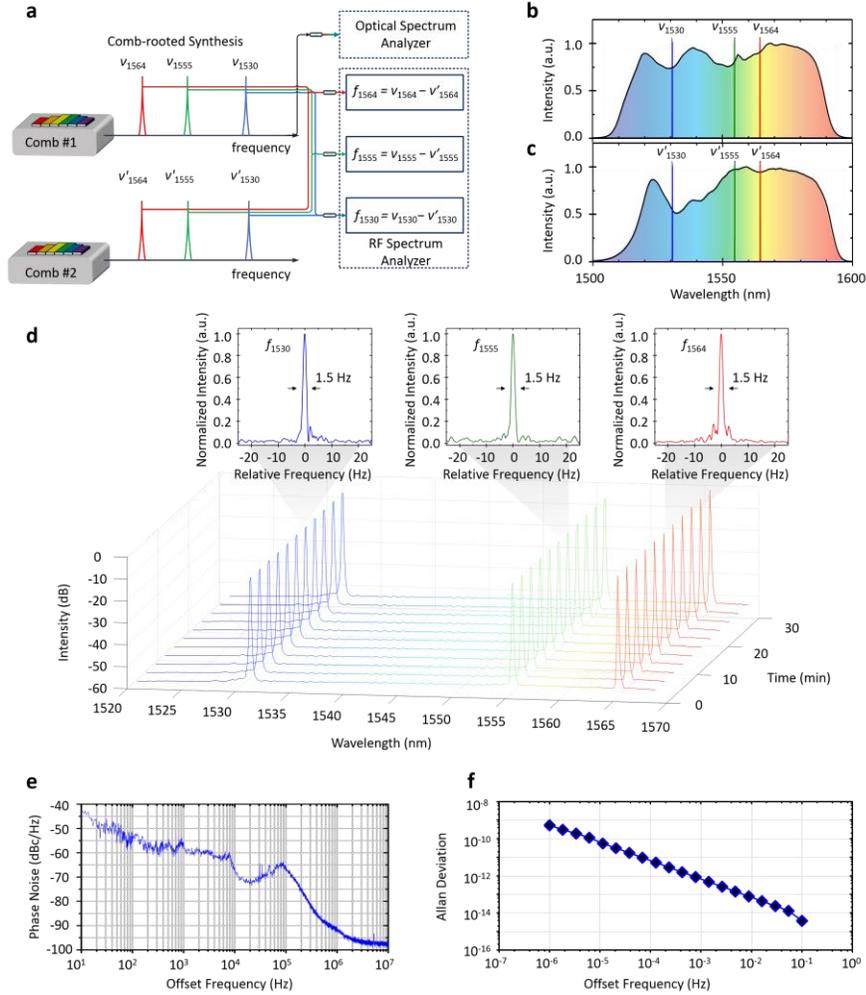

**Fig. 4. Out-of-loop test results.** (**a**) Linewidth evaluation by employing two different synthesizers; $\nu$ and $\nu'$ represent the optical frequencies from a main synthesizer (Comb #1) and an auxiliary synthesizer (Comb #2), respectively. (**b**) Optical spectrum of Comb *#1* and its three wavelength channels. (**c**) Optical spectrum of Comb *#2* and its three channels. (**d**) Out-of-loop test results of three channels with linewidth profiles monitored using an RF spectrum analyzer (RBW: 1 Hz). (**e**) Representative phase noise spectrum of RF beats (RBW: 1 Hz) and (**f**) Calculated Allan deviation of RF beats.